\documentstyle[prl,preprint,aps]{revtex} 
\input epsf
\begin{document} 
\title{Coexistence Curve Singularities at Critical End Points}

\author{Nigel B. Wilding} 

\address{Institut f\"{u}r Physik, Johannes Gutenberg Universit\"{a}t,
\\ Staudinger Weg 7, D-55099 Mainz, Germany.}

\tighten

\maketitle 
\begin{abstract}

We report an extensive Monte Carlo study of critical end point behaviour in a
symmetrical binary fluid mixture.  On the basis of general scaling arguments,
singular behaviour is predicted in the diameter of the liquid-gas
coexistence curve as the critical end point is approached.  The simulation
results show clear evidence for this singularity, as well as confirming a
previously predicted singularity in the coexistence chemical potential.  Both
singularities should be detectable experimentally.

\end{abstract}
\pacs{05.70.Jk,64.70.Ja,64.70.Fx,64.60.Fr}
\nopagebreak

A critical end point occurs when a line of second order phase transitions
intersects and is truncated by a first order phase boundary delimiting a new
noncritical phase.  Critical end points are common features in the phase
diagrams of a diverse range of physical systems, notably binary fluid
mixtures, superfluids, binary alloys, liquid crystals, certain ferromagnets
and ferroelectrics etc.  Perhaps the simplest of these is the binary fluid
mixture, for which the phase diagram is spanned by three thermodynamic
fields ($T,\mu,h$), where $T$ is the temperature, $\mu$ is a chemical
potential, and $h$ is an ordering field coupling to the relative concentrations of
the two fluid components.  In the region $h=0$, two fluid phases $\beta$ and
$\gamma$ coexist.  By tuning $T$ and $\mu$, however, one finds a critical
`$\lambda$-line', $T_c(\mu)$, where both phases merge into a single
$\beta\gamma$ phase.  This $\lambda$-line meets the first order line of
liquid-gas transitions $\mu_\sigma(T)$ at a critical end point
($T_e,\mu_e$), see fig 1.  For $T<T_e$, the phase boundary $\mu_\sigma(T)$
constitutes a triple line along which the fluid phases $\beta$ and $\gamma$
coexist with the gas phase $\alpha$, while for $T>T_e$, $\mu_\sigma(T)$
defines the region where the $\beta\gamma$ and $\alpha$ phases coexist. 
Precisely at the critical end point the critical mixture of $\beta$ and
$\gamma$ phases coexists with the gas phase.  Since the gas phase is
noncritical, it is commonly referred to as a ``spectator'' phase. 

Despite their ubiquity, it has only quite recently been pointed out that
critical end points should exhibit novel properties beyond those observable
on the critical line.  Using phenomenological scaling and thermodynamic
arguments, Fisher and coworkers \cite{FISHER1} predicted that the singular
behaviour at the critical end point also engenders new singularities in the
first order phase boundary itself.  Additionally, new universal amplitude
ratios were proposed for the shape of this boundary, as well as for the
noncritical surface tensions near the critical end point \cite{FISHER2}. 
These predictions were subsequently corroborated by analytical calculations
on extended spherical models \cite{BARBOSA}.  To date, however, empirical
support from physically realistic systems has been scarce.  While some
experimental work on surface amplitude ratios has been reported \cite{LAW},
no attention seems to have been given to the bulk coexistence properties and
the question of the predicted singularity in the spectator phase boundary. 
There is a similar dearth of simulation work on the subject, and we know of
no detailed numerical studies of critical end point behaviour, either in
lattice or continuum models. 

In this letter we move to remedy this situation by providing the first
simulation evidence for singular behaviour in the first order phase boundary
close to a critical end point.  The key features of our results are as
follows.  We consider a classical binary fluid within the grand canonical
ensemble.  We review the scaling arguments of Fisher and coworkers and
show that in addition to the previously predicted singularity in
$\mu_\sigma(T)$, they also imply singular behaviour in the diameter of the
liquid-gas coexistence curve at the critical end point.  We test these
predictions using extensive Monte Carlo simulations of a symmetrical
Lennard-Jones binary mixture, making full use of modern sampling methods,
histogram extrapolation techniques and finite-size scaling analyses.  The
results provide remarkably clear signatures of divergences in the
appropriate temperature derivatives of the coexistence diameter and the
phase boundary chemical potential. 

We shall focus attention on the liquid-gas coexistence region in the
vicinity of a critical end point.  Following Fisher and Barbosa
\cite{FISHER1}, coexistence is prescribed by the equality of the Gibbs free
energy $G=-k_BT\ln {\cal Z}$ in the gas and liquid phases i.e. 
$G_g(\mu_\sigma(T),T,h)=G_l(\mu_\sigma(T),T,h)$.  Since the gas spectator
phase is noncritical, its free energy is analytic at the end point and thus
may be expanded as

\begin{equation}
G_g(\mu,T,h)=G_e+G_1^g\Delta\mu+G_2^g t+G_3^g h +G_4^g\Delta \mu^2+\cdots
\label{eq:Ggas}
\end{equation}
where $t\equiv(T-T_e)/T_e$, $\Delta\mu\equiv\mu-\mu_e<0.$
The liquid phase on the other hand is critical and therefore contains both an
analytic (background) and a singular contribution to the free energy

\begin{equation}
G_l(\mu,T,h)=G_0(\mu,T,h)-|\tau|^{2-\alpha}{\cal G}_{\pm}(\hat{h}|\tau|^{-\Delta})
\label{eq:Gliq}
\end{equation}
where $G_0$ is the analytic part, while ${\cal G}_\pm(y)$ is a universal
scaling function whose two branches $\pm$ must satisfy matching conditions
as $y\to\pm\infty$.  The quantities $\tau(\mu,T,h)$ and $\hat{h}(\mu,T,h)$
are both scaling fields that measure deviations from criticality, and
comprise linear combinations of $\Delta\mu,t$ and $h$.  $\alpha$ and
$\Delta$ are respectively the specific heat and gap exponents associated
with the $\lambda$-line. 

Expanding the critical free energy in $\Delta \mu, t$ and $h$, and setting
$G_g(\mu_\sigma(T),T,h)= G_l(\mu_\sigma(T),T,h)$, then yields \cite{FISHER1} in
the region $h=0$

\begin{equation}
\mu_\sigma(T)-\mu_o(T)\approx -X_\pm|t|^{2-\alpha}, \hspace{5mm} T\to
T_e\pm
\end{equation}
with $\mu_o(T)=\mu_e+g_1t+\cdots$, analytic, $X_\pm>0$. If $\alpha>0$, this in turn
implies a divergence in the {\em curvature} of the spectator phase boundary

\begin{equation}
\frac{d^2\mu_\sigma}{dT^2}\approx -\tilde{X}_\pm|t|^{-\alpha}  
\label{eq:csing}
\end{equation}
where the amplitude ratio $\tilde{X}_+/\tilde{X}_-$ is expected to be universal \cite{FISHER1}.

Let us now consider the behaviour of the coexistence density in the
neighbourhood of the critical end point.  Specifically, we shall examine the
temperature dependence of the coexistence diameter in the symmetry plane $h=0$,
defined by

\begin{equation}
\rho_d(T)\equiv\frac{1}{2}[\rho_g(\mu_\sigma(T))+\rho_l(\mu_\sigma(T))]
\end{equation}
which is simply obtained from the coexistence free energy as
\begin{equation}
\rho_d(T)=-\frac{1}{V}\left ( \frac{\partial G(\mu_\sigma(T),T)}{\partial\mu}\right )
\label{eq:rho_d}
\end{equation}
where $G(\mu_\sigma(T),T)=[G_g(\mu_\sigma(T),T)+G_l(\mu_\sigma(T),T)]/2$.
Appealing to eqs.~\ref{eq:Ggas} and \ref{eq:Gliq}, one then finds

\begin{eqnarray}
\rho_d(T)  & = &-U_\pm|\tau|^{1-\alpha}-V_\pm|\tau|^\beta \nonumber\\
 & & + \hspace{0.7mm} \mbox{terms analytic at} \hspace{1mm} T_e
\label{eq:rhod}
\end{eqnarray}

This singularity is of the same form as the overall density singularity
\cite{ANISIMOV} on the critical line $T_c(\mu)$, which for binary fluids
with short ranged interactions is expected to be Ising-like.  For the
symmetrical binary fluid studied in the present work, one finds that the
amplitude $V_\pm=0$ in eq.~\ref{eq:rhod}.  The diameter derivative then
exhibits a specific heat like divergence:

\begin{equation}
\frac{d\rho_d(T)}{dT} \approx -\tilde{U}_\pm|t|^{-\alpha},
\label{eq:dsing}
\end{equation}
where we have used $|\tau|=|t|[1+O(|t|^{1-\alpha})]$ {\em along} the
coexistence curve.  Since this divergence occurs in the first derivative of
$\rho_d(T)$, it is in principle more readily visible than that in the second
derivative of $\mu_\sigma(T)$, cf.  eq.~\ref{eq:csing}.  As we shall now
show, however, clear signatures of both divergences are readily demonstrable
by Monte-Carlo simulation \cite{WILDING2}. 

The simulations described here were performed for a symmetrical binary fluid
model using a Metropolis algorithm within the grand canonical ensemble (GCE)
\cite{ALLEN}.  The fluid is assumed to be contained in (periodic) volume
$V=L^3$, with grand-canonical partition function

\begin{equation}
\label{eq:bigzdef}
{\cal Z}_{L}  = \sum _{N_1=0}^{\infty }\sum _{N_2=0}^{\infty }\prod _{i=1}^{N} \left\{\int d\vec{r}_i\right\}
e^{ \left[ \mu N
-\Phi (\{ \vec{r} \} ) + h(N_1-N_2) \right]}
\end{equation}
where $\Phi=\sum_{i<j}\phi(r_{ij})$ is the total configurational energy,
$\mu$ is the chemical potential, and $h$ is the ordering field 
(all in units of $k_BT$).  $N=N_1+N_2$ is the total number of particles of types $1$ and
$2$.  The interaction potential between particles $i$ and $j$ was assigned
the familiar Lennard--Jones (LJ) form

\begin{equation} 
\phi(r_{ij})=4\epsilon_{mn}[(\sigma/r_{ij})^{12}-(\sigma/r_{ij})^6],
\label{eq:LJdef} 
\end{equation} 
where $\sigma$ is a parameter which serves to set the interaction range,
while $\epsilon_{mn}$ measures the well-depth for interactions between
particles of types $m$ and $n$.  In common with most other simulations of
Lennard-Jones systems, the potential was cutoff at a radius $r_c=2.5\sigma$
to reduce the computational effort. 

An Ising model type symmetry was imposed on the model by choosing
$\epsilon_{11}=\epsilon_{22}=\epsilon>0$.  This choice endows the system with
energetic invariance under $h\to -h$ and ensures that the critical end point
lies in the symmetry plane $h=0$.  A further parameter
$\epsilon_{12}=\delta$ was used to control interactions between unlike
particles.  The phase diagram of the model in the surface $h=0$ is then
uniquely parameterised by the ratio $\delta/\epsilon$.  Choosing
$\delta/\epsilon\lesssim 1$ yields a phase diagram having a critical end
point temperature $T_e\ll T^{lg}_c$ and density $\rho_e\gg\rho^{lg}_c$,
where $T^{lg}_c$ and $\rho^{lg}_c$ are the liquid-gas critical temperature
and density respectively.  Choosing a smaller value of $\delta/\epsilon$,
however, moves the end point towards the liquid-gas critical point, into
which it merges for a certain sufficiently small $\delta/\epsilon$, forming
a tricritical point.  Empirically we find that the phase diagram is rather
sensitive to the choice of $\delta/\epsilon$.  Thus for
$\delta/\epsilon\approx 0.6$, we find a tricritical point, while for
$\delta/\epsilon=0.75$ there is a critical end point having $\rho_e\approx
2.3\rho^{lg}_c$.  In the present work, all simulations were performed with
$\delta/\epsilon=0.7$, which yields critical end point parameters
$T_e\approx 0.93 T^{lg}_c$, $\rho_e\approx 1.75\rho^{lg}_c$.  This
temperature is sufficiently small compared to $T^{lg}_c$ that critical
density fluctuations (which might otherwise obscure the end point behaviour)
may safely be neglected, while at the same time $\rho_e$ is not so large as
to hinder particle insertions. 

Although use of conventional GCE simulations to study liquid-gas phase
coexistence presents no great practical difficulties when $T\lesssim T_c^{lg}$
\cite{WILDING1}, investigations of the strongly first order regime $T\ll T_c^{lg}$
are rendered extremely problematic by the large free energy barrier separating
the coexisting phases.  This leads to metastability effects and protracted
correlation times.  To circumvent this difficulty we employed the
multicanonical preweighting method \cite{BERG}, which encourages the
simulation to sample the interfacial configurations of intrinsically low
probability.  This is achieved by incorporating a suitably chosen weight function in
the Monte-Carlo update probabilities.  The weights are subsequently folded out
from the sampled distributions to yield the correct Boltzmann distributed
quantities.  Further details of the implementation of this technique as well
as a method for determining a suitable preweighting function are given
elsewhere \cite{WILDING1,WILDING2}.

In the course of the simulations, three systems sizes of volume
$V=(7.5\sigma)^3$, $V=(10\sigma)^3$ and $V=(12.5\sigma)^3$ were studied,
corresponding to average particle numbers of $N\approx 250$, $N\approx 600$
and $N\approx 1200$ respectively at the critical end point (whose location
we discuss below).  Following equilibration, runs comprising up to $6\times
10^9$ MCS \cite{NOTE2} were performed and the density $\rho=N/V$, energy
density $u=\Phi/V$ and number difference order parameter $m=(N_1-N_2)/V$
were sampled approximately every $10^4$ MCS.  Attention was focused on the
finite-size distributions $p_L(\rho)$ and $p_L(m)$.  Precisely on the
liquid-gas coexistence curve, the density distribution $p_L(\rho)$ is (to
within corrections exponentially small in $L$) double peaked with equal
weight in both peaks \cite{BORGS}.  For a given simulation temperature, this
`equal weight' criterion can be used to determine the coexistence chemical
potential to high accuracy.  Simulations were carried out for each $L$ at
several (typically $5$) temperatures along the coexistence curve, and
histogram reweighting \cite{FERRENBERG} was used to interpolate between
simulation points and to aid the precise location of the coexistence
chemical potential \cite{WILDING1}.  The position of the critical end point
itself was estimated using finite-size scaling techniques in the standard
manner \cite{BINDER}, by studying the scaling of the fourth order cumulant
ratio $U_L=1-3\langle m^4\rangle/\langle m^2\rangle^2$ for $p_L(m)$ as a
function of $T$ and $L$ along the liquid branch of the coexistence curve.  A
cumulant intersection \cite{WILDING2} implying critical scale invariance was
obtained at $\tilde{T}_e=0.958(2), \rho_e=0.587(5)$, where $\tilde
T=k_BT/\epsilon$.  Further points on the $\lambda$-line away from the
critical end point were also determined using the same method.  Related
finite-size scaling techniques, this time focusing on the density-like
ordering operator \cite{WILDING1,BRUCE1} were utilised to locate the
liquid-gas critical point.  This yielded the estimates
$\tilde{T}_c^{lg}=1.024(2), \rho_c^{lg}=0.327(2)$. 

Figure~\ref{fig:cxcurve} shows the estimated coexistence liquid and gas
densities as a function of temperature, determined as the peak densities of
$p_L(\rho)$.  Also shown is the measured locus of the $\lambda$-line and the
position of the liquid-gas critical point.  Clearly a pronounced `kink' is
discernible in the liquid-branch density in the vicinity of the critical end
point.  The gas branch, on the other hand displays no such kink due to the
analyticity of $G_g(\mu,T)$ at $T_e$.  To probe more closely the behaviour
of the coexistence density, we plot in figure~\ref{fig:diam}(a) the diameter
derivative $-d\rho_d/dT$, for the three system sizes studied.  The data
exhibit a clear peak close to $T_e$, that narrows and grows with increasing
system size.  Very similar behaviour is also observed in the curvature of
the spectator phase boundary $-d^2\mu_\sigma/ dT^2$, see figure
~\ref{fig:diam}(b).  These peaks constitute, we believe, the
finite-size-rounded forms of the divergences eqs.~\ref{eq:dsing} and
\ref{eq:csing}.  On the basis of finite-size scaling theory \cite{BINDER},
the peaks are expected to grow in height like $L^{\alpha/\nu}$, with $\nu$
the correlation length exponent.  Unfortunately it is not generally feasible
to extract estimates of $\alpha/\nu$ in this way (even for simulations of
lattice Ising models), because to do so necessitates an accurate measurement
of the analytic background, for which the present system sizes are much too
small.  Nevertheless, the correspondence of the peak position with the
independently estimated value of $\tilde{T}_e$, as well as the narrowing and
growth of the peak with increasing $L$ constitutes strong evidence
supporting the existence of the predicted singularities. 

In summary we have employed advanced Monte-Carlo simulation techniques to
study the first order phase boundary near the critical end point of a
continuum binary fluid model.  The results provide the first empirical
evidence for singularities in the phase boundary and the coexistence curve
diameter.  We expect that similar effects should be experimentally
observable (not only in binary fluids), and that due to the absence of
finite-size rounding they should be even more marked than observed here. 
Moreover, as we have shown, for asymmetrical systems such as real binary
mixtures, the coexistence diameter is expected to manifest a much stronger
singularity than occurs in the present symmetrical model. This should
therefore be particularly conspicuous. 

The author thanks K.  Binder and D.P.  Landau for stimulating discussions. 
Helpful correspondence with A.D.  Bruce, M.E.  Fisher, M.  Krech and M. 
M\"{u}ller is also gratefully acknowledged.   This work was supported   
by BMBF project number 03N8008 C.

\newpage
\begin{figure}[h]
\setlength{\epsfxsize}{9.0cm}
\centerline{\mbox{\epsffile{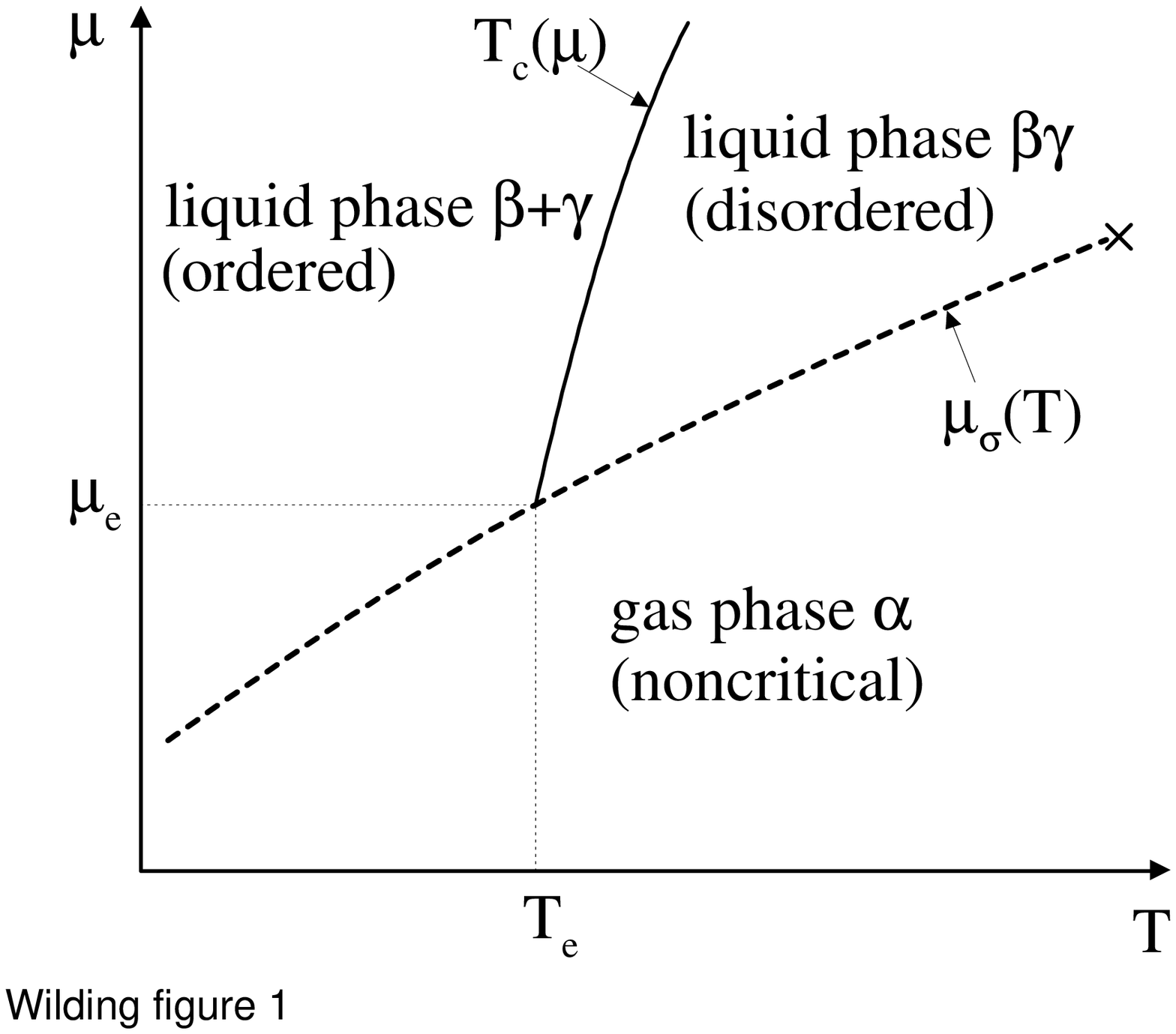}}} 

\caption{Schematic phase diagram of a binary fluid in the coexistence
surface $h=0$.  The broken line $\mu_\sigma(T)$ is the first order
liquid-gas phase boundary terminating at the critical point (cross).  The
full line is the critical line of second order transitions $T_c(\mu)$
separating the demixed phases $\beta$+$\gamma$, from the mixed phase
$\beta\gamma$.  The two lines intersect at the critical end point.}

\label{fig:meth}
\end{figure}

\begin{figure}[h]
\setlength{\epsfxsize}{9.0cm}
\centerline{\mbox{\epsffile{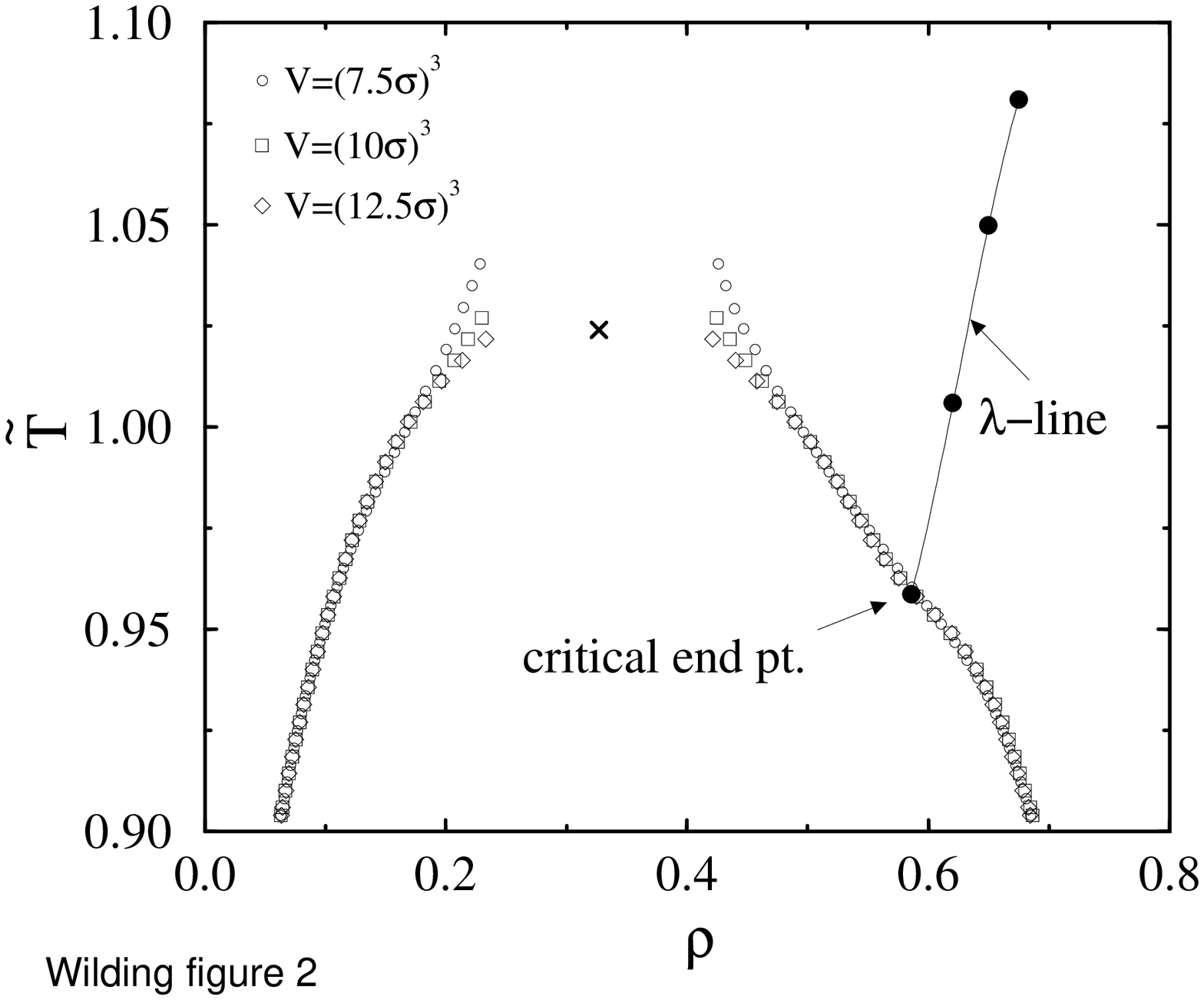}}} 

\caption{The peak densities corresponding to the coexistence form of
$p_L(\rho)$ for the three systems sizes studied, plotted as a function of
the temperature.  Also shown is the estimated locus of the $\lambda$-line
(circles) and the liquid-gas critical point (cross).  Statistical errors do
not exceed the symbol sizes.}

\label{fig:cxcurve}
\end{figure}
\vspace{10cm}
\begin{figure}[h]
\setlength{\epsfxsize}{9.0cm}
\centerline{\mbox{\epsffile{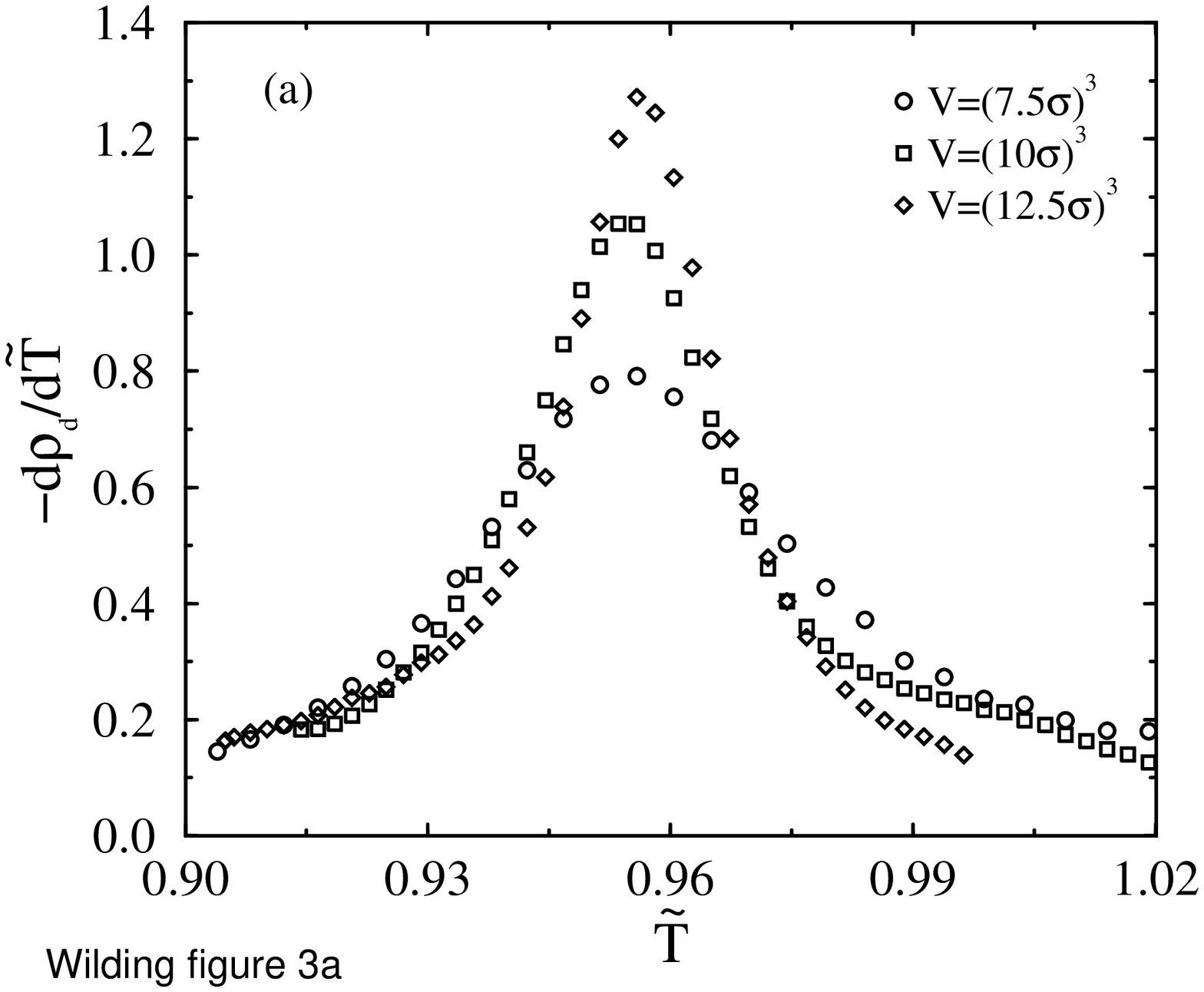}}} 
\centerline{\mbox{\epsffile{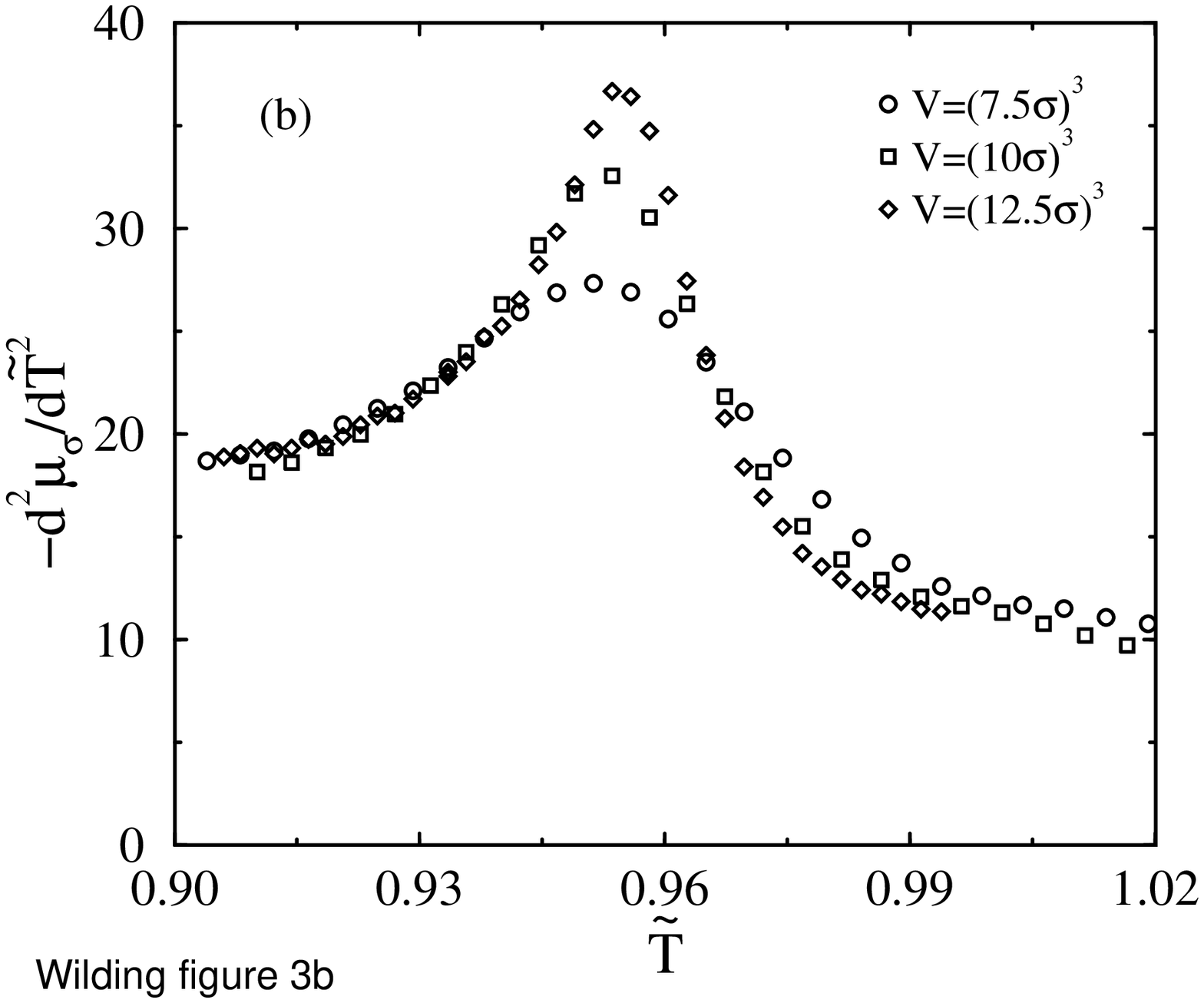}}} 

\caption{{\bf (a)} The numerical temperature derivative of the measured
coexistence diameter $-d\rho_d/dT^\star$ in the vicinity of the critical end point
temperature. {\bf (b)} The measured curvature of the phase boundary, $-d^2\mu_\sigma/dT^2$,
in the vicinity of the critical end point temperature.}

\label{fig:diam}
\end{figure}

\end{document}